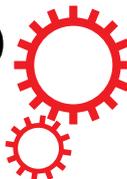

# SCIENTIFIC REPORTS

**OPEN** 

# Origin of long-lived quantum coherence and excitation dynamics in pigment-protein complexes

Zhedong Zhang[1,2] & Jin Wang[1,3,4]

We explore the mechanism for the long-lived quantum coherence by considering the discrete phonon modes: these vibrational modes effectively weaken the exciton-environment interaction, due to the new composite (polaron) formed by excitons and vibrons. This subsequently demonstrates the role of vibrational coherence which greatly contributes to long-lived feature of the excitonic coherence that has been observed in femtosecond experiments. The estimation of the timescale of coherence elongated by vibrational modes is given in an analytical manner. To test the validity of our theory, we study the pigment-protein complex in detail by exploring the energy transfer and coherence dynamics. The ground-state vibrational coherence generated by incoherent radiations is shown to be long-survived and is demonstrated to be significant in promoting the excitation energy transfer. This is attributed to the nonequilibriumness of the system caused by the detailed-balance-breaking, which funnels the downhill migration of excitons.

Recently the widespread interest in exploring the quantum nature in solar cells and the photosynthetic process has been triggered by experimental investigations of excitonic dynamics in light-harvesting and Fenna-Matthews-Olson (FMO) complexes[1–3]. The transport of excitation energy in the antenna is remarkably fast and efficient, usually with quantum yields close to 100%[4]. Even with the knowledge of electronic structure in antenna and the advances in spectroscopy that uncovered the long-lived quantum coherence in noisy environment[5,6], the full understanding of the role of coherence and mechanism of excitation energy transfer has still remained mysterious.

Conventionally, either in Förster theory or the advanced models including dephasing[7–10], the excitonic energy transfer is considered in adiabatic framework under Born-Oppenheimer approximation[11] where only the degrees of freedoms of the electrons are included. Although the models under this approximation were somehow successful in describing the population dynamics of excitons in a quasi-classical way, they show their failure on explaining the long-lived coherence oscillations[7,12] observed in 2D femtosecond electronic spectroscopy[13–17]. In fact, some discrete intramolecular vibrations are of comparable time scale with the relaxation of exciton, which subsequently leads to the breakdown of adiabatic approximation[18–21]. Thus this phonon dynamics often has a crucial effect on the energy transport when the energy quanta of vibrational modes are in resonance with the energy splitting of excitons[22–25]. Here the non-adiabaticity explicitly refers to the case including the phonon dynamics owing to the comparable relaxation of phonons with the excitons. Such effect is not considered in the adiabatic regime, since the phonons can be averaged out because of the much faster relaxation of phonons than that of excitons. This is in contrast with the non-adiabatic case where phonons cannot be easily averaged out due to their slow relaxation. Therefore the exciton-vibraiton interaction have to be explicitly considered in this case. The persistence of coherences originating from the vibrational modes has been evidently observed in recent experiment[26].

In this article we develop an effective theory to explore in a general scenario the underlying mechanism for long-lived quantum coherence, while other investigations *numerically* show the existence of such effect without illustrating the mechanism at rigorous level[18,24]. The bare electron is surrounded by discrete vibrational modes whose dynamics will be taken into account, due to their strong coupling and subsequently comparable relaxation timescale with electrons. As a result, the new composite called polaron emerges, resulting in the suppression of system-reservoir coupling. Intuitively this configuration gives rise to the screening of the exciton-bath interaction, in analogy to the Yukawa potential $\sim \frac{e^{-\lambda r}}{r}$ between electrons in the interacting electron gas[27]. This leads to much longer survival of quantum coherence than that with bare excitons only. Our general theory, on the other hand, suggests a physical mechanism for slowing down the dynamical decoherence, which is potentially applicable in quantum computation. Our motivation is to understand the role of vibrational coherence on the long-lived coherence and also the efficient energy transfer measured in recent experiments[14–16]. Our general effective theory is verified in the pigment-protein complex by uncovering the long-lived electronic and excitonic coherences caused by vibrational coherence. Moreover in such a system, the ground-state coherence contributed by incoherent environment is found to give rise to the considerable enhancement of excitation energy transfer. This, however, fails to be predicted by the previous model without incoherent radiations.

[1]Department of Physics and Astronomy, SUNY Stony Brook, Stony Brook, NY 11794, USA. [2]Department of Chemistry, University of California Irvine, Irvine, CA 92697, USA. [3]Department of Chemistry, SUNY Stony Brook, Stony Brook, NY 11794, USA. [4]State Key Laboratory of Electroanalytical Chemistry, Changchun Institute for Applied Chemistry, Chinese, Academy of Sciences, Changchun, Jilin 130022, P. R. China. Correspondence and requests for materials should be addressed to J.W. (email: jin.wang.1@stonybrook.edu)

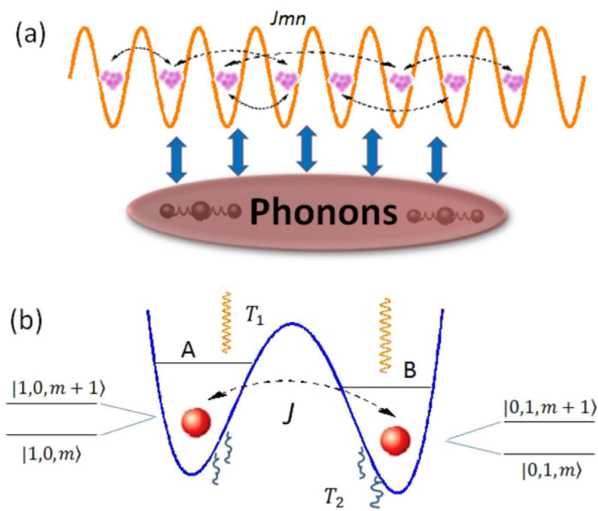

**Figure 1.** The schematic of (**a**) our model describing exciton-phonon interaction in Eq. (1) and (**b**) pigment-protein complex. In (**a**) each well represents a site and the whole system interacts with phonon environment; In (**b**) the excitons in pigments couple to a vibrational mode and the radiation energy with temperature $T_1$ is absorbed by such joint system and then dissipated into the noisy protein environment with temperature $T_2$.

## Model

For the general consideration of the coupling of molecular excitations to the motions of molecular vibrations, we introduce the $N$-site fermionic system coupled to the phonons, as shown in Fig. 1(a). The subsequent Hamiltonian is

$$H = \sum_{n=1}^{N} \varepsilon_n a_n^\dagger a_n + \sum_{n<m} J_{nm}(a_n^\dagger a_m + a_m^\dagger a_n) + \sum_{\mathbf{q},s} \hbar \omega_{\mathbf{q}s} b_{\mathbf{q}s}^\dagger b_{\mathbf{q}s}$$
$$+ \sum_{\mathbf{q},s} \sum_{n=1}^{N} f_{\mathbf{q}s} \hbar \omega_{\mathbf{q}s} \gamma_n a_n^\dagger a_n (b_{\mathbf{q}s} + b_{-\mathbf{q},s}^\dagger)$$

(1)

where the 1st, 2nd and 3rd terms of $H$ represent the onsite energy of excitons, electronic coupling between different sites and free Hamiltonian of phonons, respectively. In the last term $f_{\mathbf{q}s}$ governs the strength of electron-phonon interaction and $\gamma_n$ quantifies the diagonal disorder of the system, leading to the renormalization of the on-site energies as shown later. $\omega_{\mathbf{q}s}$ stands for the dispersion relation of phonons as a function of wave vector $\mathbf{q}$ and $s$ denotes the phonon polarization. $a_n$ and $b_{\mathbf{q}s}$ are the fermionic and bosonic annihilation operators of excitons and phonons, respectively. To take into account the effect of discrete vibrational modes, we need to reach the effective theory, by applying the polaron transform to the entire system with the generating funtion $S = \sum_{m=1}^{N} \sum_{\mathbf{q}s} f_{\mathbf{q}s} \gamma_m a_m^\dagger a_m (b_{-\mathbf{q}s}^\dagger - b_{\mathbf{q}s})$. Such a polaron transform has been developed in condensed matter physics[27,28] and adopted to molecular aggregates[21,29–32], in order to explore the effect of strong exciton-bath couplings on the exciton dynamics immersing in phonon environment[29,30]. Our idea in this study gives a different scenario in that those discrete vibrational modes in quasi-resonance with excitonic energy gap are separated from others. This leads to the strong coupling of these modes to excitons and therefore their dynamics should be simultaneously considered as well. The treatment of this phonon dynamics was lacking before. Because of the off-resonance with excitons, the remaining phonon modes weakly interact with the excitons. This configuration results in the effective Hamiltonian $\widetilde{H} = H_S + H_{ph} + H_{int}$, illustrated in Fig. 2(b) and

$$\begin{aligned} H_S &= \sum_{i=1}^{N} \bar{\varepsilon}_i a_i^\dagger a_i + 2V_0 \sum_{i<j} \gamma_i \gamma_j a_i^\dagger a_j^\dagger a_i a_j + \sum_n \hbar\omega_n d_n^\dagger d_n \\ &\quad + \sum_{i<j} J_{ij} \left[ \prod_n e^{\lambda_n(\gamma_i - \gamma_j)(d_n^\dagger - d_n)} a_i^\dagger a_j + \text{h.c.} \right] \\ H_{int} &= \sum_{\mathbf{q},s} \sum_{i<j} (\gamma_i - \gamma_j) J_{ij} \prod_m e^{-\lambda_m^2(\gamma_i - \gamma_j)^2/2} f_{\mathbf{q}s} \\ &\quad \times \left( a_i^\dagger a_j \prod_n \chi_{ij}^n - a_j^\dagger a_i \prod_n \chi_{ij}^{n,\dagger} \right) (b_{-\mathbf{q},s}^\dagger - b_{\mathbf{q}s}), \\ H_{ph} &= \sum_{\mathbf{q},s}{}' \hbar\omega_{\mathbf{q}s} b_{\mathbf{q}s}^\dagger b_{\mathbf{q}s} \end{aligned} \quad (2)$$

where $\bar{\varepsilon}_i = \varepsilon_i - \gamma_i^2 V_0$ is the renormalized onsite energy in the 1st term of $H_S$ and $V_0 = \sum_{\mathbf{q},s} f_{\mathbf{q}s}^2 \hbar\omega_{\mathbf{q}s}$. The 2nd and 3rd terms are the exciton-exciton interaction (intermediated by phonons) and the free Hamiltonian of discrete vibrational modes, respectively. The last term in $H_S$ quantifies the electronic coupling renormalized by exciton-vibrational (discrete phonon modes) coupling. The exciton-phonon interaction strength $f_{\mathbf{q}s}$ is replaced by $\lambda_n$ where $\lambda_n = f_{\mathbf{q}'s'}$ for those discrete modes. $|\lambda_n \gamma_i|$ refers to the Huang-Rhys or Frank-Condon factor[33] which quantifies the overlap between the vibrational wavefunctions. The vibrational operator is $\chi_{ij}^n = e^{\lambda_n(\gamma_i - \gamma_j)d_n^\dagger} e^{-\lambda_n(\gamma_i - \gamma_j)d_n}$. Those discrete vibrational modes are denoted by the operator $d_n$'s. $H_{int}$ above is obtained up to the 1st order of $f_{\mathbf{q}s}$ and it describes the coupling of excitons to the remaining phonon modes, other than the discrete phonon modes. Equation (2) shows that the effective coupling strength between exciton and phonon is renormalized by the polaron effect, which leads to the weak interaction of the new composite (exciton + vibron) with the environmental modes, as will be illustrated in details in the section of Results. $H_{ph}$ denotes the Hamiltonian of phonon environment.

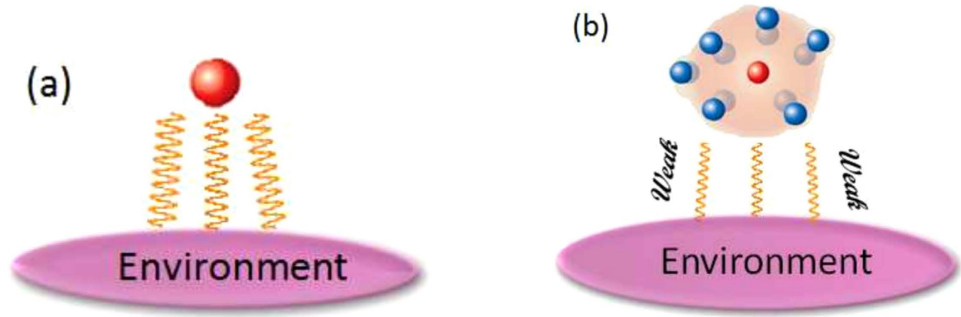

**Figure 2. The schematic for exciton-environment interaction.** (**a**) The coupling between bare exciton and environment; (**b**) The exciton-vibration coupling forms a new composite called polaron. In (**b**) the bare exciton (red) is surrounded by a cloud consisting of discrete vibrational modes (blue), which leads to the suppression of the coupling strength between exciton and environment.

## Results

**General mechanism for long-lived quantum coherence.** Based on Eq. (2), the discrete phonon modes are glued to excitons and the whole system forms the polarons, which weakly couple to phonon environment consisting of quasi-continuous phonon modes. This results in the renormalization of the effective coupling strength between exciton and phonon, as illustrated in Fig. 2. To elucidate this issue in detail, we consider the phonon dynamics being restricted to the space spanned by $|\{m_n\}\rangle$, $|\{m_n + 1\}\rangle$, where $m_n = \langle\{m_n\}|d_n^\dagger d_n|\{m_n\}\rangle$ represents the occupation number of phonons on each vibrational mode. This is due to the quasi-resonance in energy between the exciton transition and vibrational modes. In most circumstance, the molecular vibrations are always excited from the ground state so that only the eigenstates $|\{0\}\rangle$, $|\{1\}\rangle$ are included. Therefore by taking into account the matrix elements of the vibrational operator $\chi_{ij}^n$: $\langle\{m_n\}|\chi_{ij}^n|\{m_n\}\rangle = L_{m_n}(\lambda_n^2(\gamma_i - \gamma_j)^2)$, $\langle\{m_n\}|\chi_{ij}^n|\{m_n + 1\}\rangle = -(\gamma_i - \gamma_j)\lambda_n\sqrt{m_n + 1} \,_1F_1(-m_n;2;\lambda_n^2(\gamma_i - \gamma_j)^2)$, which are polynomials with order ~1 of magnitude especially for the case $m_n = 0$, the effective coupling between system and phonon environment is renormalized as

$$\bar{f}_{\mathbf{q}s}^{ij} = (\gamma_i - \gamma_j) \prod_n e^{-\lambda_n^2(\gamma_i - \gamma_j)^2/2} f_{\mathbf{q}s} \quad (3)$$

which shows that the system-bath interaction is suppressed by a factor of $\prod_m e^{-\lambda_m^2(\gamma_i - \gamma_j)^2/2}$. In the framework of weak system-reservoir interaction, the typical dephasing rate $\Gamma$ is determined by the square of system-bath coupling, namely, $\Gamma \sim f_{\mathbf{q}s}^2$ [34]. This leads to the reduction of the dephasing rate induced by phonon environment, namely, $\overline{\Gamma} = (\gamma_i - \gamma_j)^2 \frac{\langle l_{ij}^2 \rangle}{\hbar^2 \langle \overline{\omega} \rangle^2} \prod_n e^{-\lambda_n^2 (\gamma_i - \gamma_j)^2} \Gamma \simeq \prod_n e^{-\lambda_n^2 (\gamma_i - \gamma_j)^2} \Gamma$ and then the typical lifetime of coherence is elongated

$$\frac{\overline{\tau}}{\tau} \simeq \prod_n \frac{e^{\lambda_n^2 (\gamma_i - \gamma_j)^2}}{(\gamma_i - \gamma_j)^2} \qquad (4)$$

Notice that in the matrix elements of $\chi_{ij}^n$, $L_n(z)$ is the Laguerre polynomial and $_1F_1(a; b; z)$ is the Hypergeometric function of order (1, 1). Equation (4) demonstrates that the lifetime of coherence of the system is *exponentially* improved by the exciton-vibration coupling. In pigment-protein complexes such as FMO systems, $|\gamma_i - \gamma_j| \sim 2$ and $\lambda_n \sim 1$ so that $\overline{\tau}_2 \gtrsim 10 \tau_2$ if $n = 1$, namely, only one discrete vibrational mode is included. Furthermore it should be noticed that the amplification factor of the typical lifetime of coherence will be $e^{\lambda^2 (\gamma_i - \gamma_j)^2 M}$ when considering $M$ discrete vibrational modes in quasi-resonance with exciton energetic gap. As a brief summary, this demonstrates the mechanism in a general scenario: some modes of molecular vibrations surrounding the exciton form the new quasiparticle called polaron which results in the screening of interaction between bare exciton and environment. Hence we suggest that the consequent weak coupling of polarons to environment is the origin of the long-lived quantum coherence. It is also worthy to point out that based on this mechanism, *the vibrational coherence generated by vibrational modes will play a significant role in understanding the long-lived excitonic coherence.*

**Pigment-protein Complex.** As an example for elucidating the general mechanism proposed in the last section, we will study in detail the coherence dynamics of the pigment-protein complex, as illustrated in Fig. 1(b). By considering a prototype dimer strongly coupled to intramolecular vibration of frequency $\omega$, the interaction between exciton and vibrational modes reads

$$H_{ex-vib}^{pp} = (\eta \hbar \omega / 2^{1/2})(|A\rangle\langle A| - |B\rangle\langle B|)(d_- + d_-^\dagger) \qquad (5)$$

for the non-adiabatic treatment and total Hamiltonian is $H_S^{pp} = H_0^{pp} + H_{vib} + H_{ex-vib}^{pp}$, where $H_0^{pp} = \varepsilon_A |A\rangle\langle A| + \varepsilon_B |B\rangle\langle B| + J(|A\rangle\langle B| + |B\rangle\langle A|)$ and $H_{vib} = \hbar \omega d_-^\dagger d_-$. $|A\rangle$ and $|B\rangle$ denote the electronic states of pigments $A$ and $B$, respectively. The minus sign in Eq. (5) is due to the relative motion between the vibrational modes[35]. $\eta$ is the Franck-Condon factor of the vibrations. The energy gap between the localized excitons is $\Delta = \varepsilon_A - \varepsilon_B > 0$.

In pigment-protein complexes, the excitons interact with both the incoherent radiation and the low-frequency fluctuations of protein (phonons), to funnel the unidirectional energy transfer. The interaction with radiation takes the dipolar form of $\mathbf{p} \cdot \mathbf{A}$ in *quantum electrodynamics*, where $\mathbf{p}$ and $\mathbf{A}$ are the exciton momentum and vector potential, respectively. The interaction then reads

$$H_{int}^{pp} = \sum_{n=A,B} \sum_{\mathbf{k},p} g_{\mathbf{k}p}(a_n + a_n^\dagger)(r_{\mathbf{k}p} + r_{-\mathbf{k},p}^\dagger) + \sum_{n=A,B} \sum_{\mathbf{q},s} f_{\mathbf{q}s} \gamma_n a_n^\dagger a_n (b_{\mathbf{q}s} + b_{-\mathbf{q},s}^\dagger) \qquad (6)$$

where $a_n$ and $a_n^\dagger$ are the annihilation and creation operators of excitons on pigments $A$ and $B$, respectively. In the Hilbert space of our model $\mathcal{H} = \mathcal{H}_{ele}^{(A)} \otimes \mathcal{H}_{ele}^{(B)} \otimes \mathcal{H}_{vib}$, the annihilation operators of excitons can be expressed as $a_A = \sigma^- \otimes 1 \otimes 1$, $a_B = \sigma^z \otimes \sigma^- \otimes 1$ to obey the fermionic commutation relation, where $\sigma^\pm$, $\sigma^z$ are Pauli matrices operating on the local eigenstates of pigments. Particularly $|A\rangle \equiv |1_A, 0_B\rangle = a_A^\dagger |0\rangle$, $|B\rangle \equiv |0_A, 1_B\rangle = a_B^\dagger |0\rangle$ where $|0\rangle$ stands for the exciton vacuum. $r_{\mathbf{k}p}$ and $b_{\mathbf{q}s}$ are the bosonic annihilation operators of the radiation and low-frequency fluctuation environments, respectively. The free Hamiltonian of the reservoirs is $H_{bath} = \sum_{\mathbf{k},p} \hbar \nu_{\mathbf{k}p} r_{\mathbf{k}p}^\dagger r_{\mathbf{k}p} + \sum_{\mathbf{q},s} \hbar \omega_{\mathbf{q}s} b_{\mathbf{q}s}^\dagger b_{\mathbf{q}s}$. Usually the low-frequency fluctuations are effectively described by the Debye spectral density $S(\omega) = (2 E_R / \pi \hbar)(\omega \omega_d / (\omega^2 + \omega_d^2))$ where $E_R$ is the so-called reorganization energy and it governs the coupling strength between system and low-energy fluctuations. $\omega_d$ refers to the Debye cut-off frequency. Owing to fast relaxation of the environments and weak system-bath interaction as pointed out in our model, the quantum master equation (QME) for the reduced density matrix of the systems can be derived by tracing out the degree of freedoms of the environments. In Liouville space the QME can be further formulated as two-component form

$$\frac{\partial}{\partial t}\begin{pmatrix} \rho_p \\ \rho_c \end{pmatrix} = \begin{pmatrix} \mathcal{L}_p & \mathcal{L}_{pc} \\ \mathcal{L}_{cp} & \mathcal{L}_c \end{pmatrix} \begin{pmatrix} \rho_p \\ \rho_c \end{pmatrix} \qquad (7)$$

where $\rho_p$ and $\rho_c$ represent the population and coherence components of the density matrix, respectively. $\mathcal{L}_{pc}$ and $\mathcal{L}_{cp}$ quanfify the *non-trivial* entanglement between the dyanmics of population and electronic coherence, beyond the secular approximation. Their forms can be directly obtained from Eq. (S20) in Supporting Information (SI) and the details are omitted here to avoid redundancy.

**Coherence dynamics and excitation energy transfer.** As is known, the excitation energy transfer is reflected by the transcient behavior of the population on pigment $B$. This is quantified by the scaled probability $\overline{P}_B(t) = P_B(t)/(P_A(t) + P_B(t))$ where $P_i(t) = \langle i| \text{Tr}_{vib}(\rho) |i\rangle = \sum_{\nu=m}^{m+1} \langle i, \nu | \rho | i, \nu \rangle$, $i = A, B$. To clarify the contribution of quantum coherence, we denote the coherence in the *localized basis* into two groups: electronic and vibronic coherences. The former and latter ones are defined as $C_{ele} = \text{Im}\langle A| \text{Tr}_{vib}(\rho) |B\rangle = \text{Im}\langle A, m | \rho | B, m \rangle + \text{Im}\langle A, m + 1 | \rho | B, m + 1 \rangle$ and $C_{vib}^{(i)} = \text{Im}\langle i, m | \rho | i, m + 1 \rangle$, $i = A, B$. The notation $\text{Tr}_{vib}$

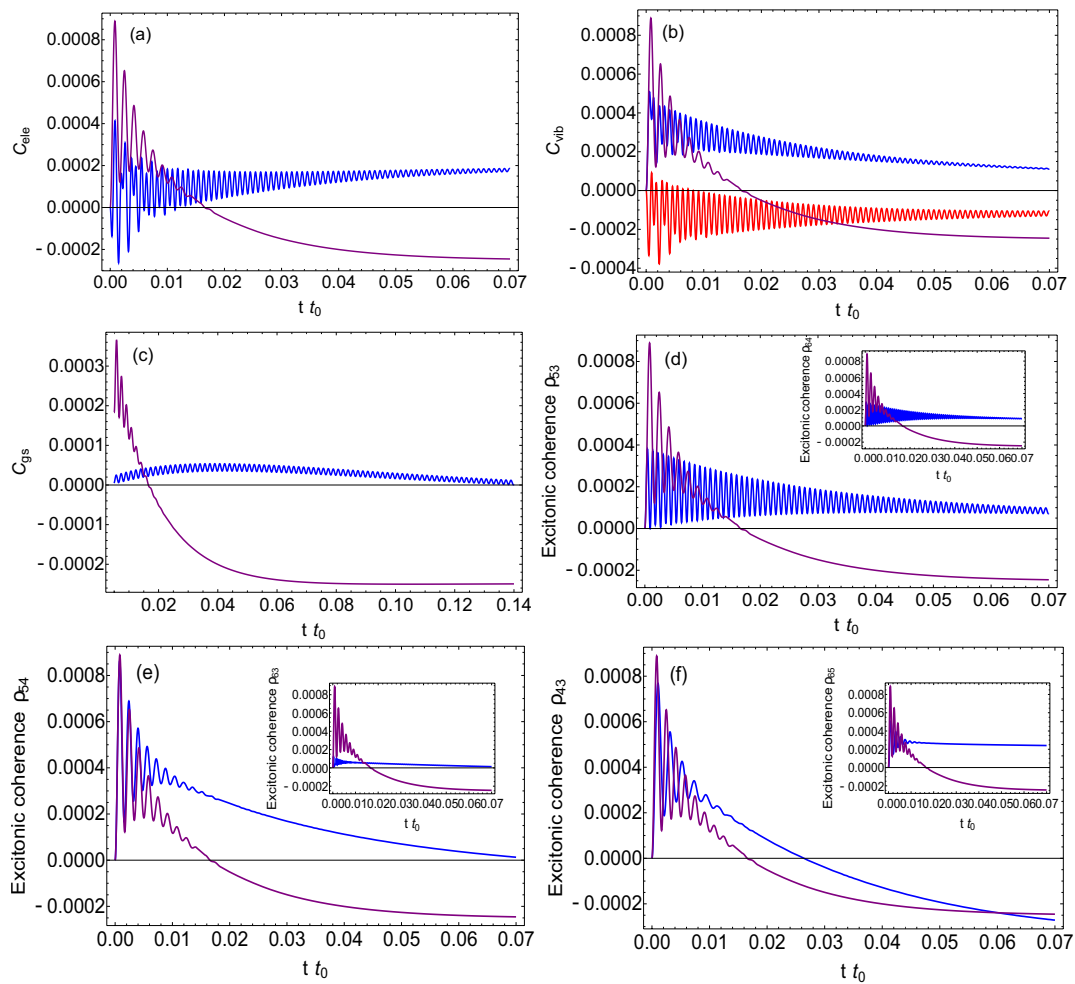

**Figure 3.** Top: Time evolution of (**a**) electronic wave packet, (**b**) excited-state vibrational wave packet and (**c**) ground-state vibrational wave packet. In (**a,c**), blue lines are for our model in non-adiabtic regime; In (**b**) the blue and red lines correspond to excited-state vibrational wave packets $\langle A,0|\rho|A,1\rangle$ and $\langle B,0|\rho|B,1\rangle$, respectively. Bottom: Time evolution of the *excitonic* coherence in delocalized basis, where blue lines correspond to non-adiabatic regime. Purple lines in (**a,b,c**) electronic and (**d,e,f**) excitonic coherences are for adiabatic regime, respectively. The parameters are: Frank-Condon factor $\eta \simeq 1$, $\Delta = 0.023\varepsilon_A$, electronic coupling $J = 0.01\varepsilon_A$, frequency of vibrational mode $\hbar\omega = 1.33\Delta$, $k_B T_1 = 0.63\varepsilon_A$, $k_B T_2 = 1.4\Delta$, Debye frequency $\hbar\omega_d = 0.7\Delta$, typical decay rate $\hbar\gamma = 0.0005\varepsilon_A$, reorganization energy $E_R = 0.23\Delta$ and $t_0 = 10\gamma^{-1}$.

means the partial trace over the degree of vibrational modes. Moreover the ground-state coherence takes the form of $C_{gs} = \text{Im}\langle 0,m|\rho|0, m+1\rangle$. As will be shown later, these vibronic coherences will dramatically affect the behavior of energy transfer and decoherence processes. In our calculations, the vibrational mode is assumed to be excited at ground state, namely, $m = 0$.

*Long-lived coherence.* To explore the effect of molecular vibrations on the decoherence process, we will study the coherence dynamics, for both the cases including exciton-vibrational coupling (non-adiabatic regime) or not (adiabatic regime). Figure 3(a) shows the dynamical behavoirs of the electronic coherences in both the adiabatic (purple) and non-adiabatic (blue) regimes. Obviously the non-adiabatic framework makes the electronic wave-packet $C_{ele}$ become long-lived in oscillation, comparing to the case in adiabatic regime. Quantitatively, the time constant of beating of the electronic wavepacket with exciton-vibron coupling is $\tau_{na} \sim 0.024t_0$ while it is $\tau_a \sim 0.005t_0$ without exciton-vibron coupling. Taking FMO complex as example $\Delta \simeq 150\,\text{cm}^{-1}$, $\omega \simeq 200\,\text{cm}^{-1}$, $J \simeq 66\,\text{cm}^{-1}$ and $\gamma \simeq 0.1\,\text{ps}^{-1}$ [19,36], thus $\tau_{na} \sim 2.2\,\text{ps}$ and $\tau_a \sim 500\,\text{fs}$. This confirms the mechanism given by Eqs (3) and (4) in our model such that the weak coupling of polaron to environment by including exciton-vibrational interaction serves as the origin for the long-lived electronic coherence. The structure of QME in Eq. (7) reveals that the environments are forbidden to generate the direct transition between vibrational states, namely, between $|f, 0\rangle$, $|f, 1\rangle$; $f = A, B$. This, in other words, will lead to the long surviving time of vibrational wave packet oscillation at excited states, as reflected in Fig. 3(b). The ground-state vibronic coherence $C_{gs}$ holds extremely long-lived oscillation of ground-state wave packet, compared to the vibrations of excited-state wave packet (the time scale for GS wave packet is $\geq 0.2t_0$ while it is $\sim 0.024t_0$ for excited-state wave packet), as shown in Fig. 3(c).

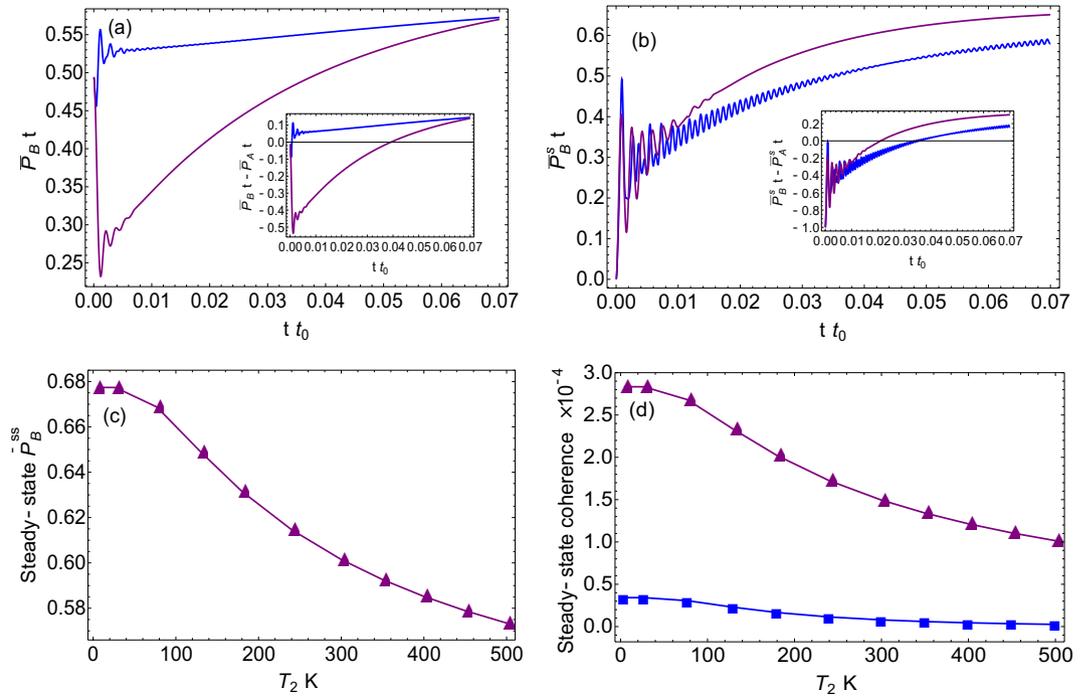

**Figure 4.** The dynamics of scaled population on pigment $B$ for (**a**) including and (**b**) NOT including the incoherent radiation environment. In both (**a**,**b**), the blue and purple curves correspond to the non-adiabatic and adiabatic regimes, respectively. (**c**) Steady-state population on pigment $B$ with respect to the temperature of low-frequency fluctuations; (**d**) Steady-state quantum coherence varies as a function of the temperature of low-frequency fluctuations. In (**d**) the purple and blue lines are for electronic (localized) and excitonic (delocalized) coherences, respectively. The parameters are the same as in Fig. 3.

Such extremely long surviving time is attributed to the fact that there is no chance for the exciton to decay when the wave packet is oscillating in the ground state.

In 2D femtosecond experiments, coherence dephasing appears as the decay of oscillations in the amplitudes of the cross peaks, which describes the superpositions of excitonic states (localized). Thereby the position $(\omega_\tau, \omega_t)$ of the cross-peak refers to the delocalized excitonic states rather than the localized ones. In such spirit, we should therefore investigate the *excitonic coherence* as shown in delocalized basis, in distinction from the electronic as well as vibronic coherences in localized basis before. From Fig. 3(d) we can see that the oscillation of the excitonic coherence is long-lived with the surviving time $\bar{\tau}_{na} \sim 0.024 t_0$ being at least ~5 times than $\bar{\tau}_a \sim 0.005 t_0$ in adiabatic regime, because of the long-lived oscillations of vibrational and electronic wave packets as shown in Fig. 3(a) and (b). This numerical evaluation of the lifetime of excitonic coherence illustrates the validity of the mechanism of long-lived coherence shown in Eqs (3) and (4), as suggested by our model. For FMO complex from green sulfur bacteria, $\bar{\tau}_{na} \sim 2.4$ ps and $\bar{\tau}_a \sim 500$ fs, which reveals the long-survived feature of coherence observed in recent experiments[14–16], even though only one vibrational mode is considered in this simplified model. In reality, the long-lived coherence is always observed by including several vibrational modes coupled to excitons and this issue will be touched in the last part of this subsection. It further turns out that the excitation energy transfer on the paths $(|A, 0\rangle; |B, 0\rangle)$ and $(|A, 1\rangle; |B, 1\rangle)$ to be coherent while on the other paths it is still incoherent, due to the short-lived oscillation of wave packet between other states, as shown in Fig. 3(e) and (f).

*Population dynamics and energy transfer.* To uncover the effect of incoherent radiation on energy transfer, we need to study the time evolution of population on pigment $B$, for both adiabatic and non-adiabatic regimes, as shown in Fig. 4, where incoherent radiation is included in 4(a) but not in 4(b). The initial conditions are: (a) $\rho(0) = |0, 0\rangle\langle 0, 0|$ for blue and $\rho(0) = |0\rangle\langle 0|$ for purple; (b) $\rho(0) = |A, 0\rangle\langle A, 0|$ for blue and $\rho(0) = |A\rangle\langle A|$ for purple. By comparing Fig. 4(a) and (b), one can conclude that the vibrational coherence, especially ground-state vibrational coherence, facilitates the excitation energy transport by including the incoherent radiations (blue line is higher than purple in Fig. 4(a)). Otherwise, it is unable to promote the energy transfer process (blue line is lower than purple in Fig. 4(b)). In particular, the incoherent environment (radiation) induces the coupling among the dynamics of excitation populations $P_i(t)$; $i = A, B$ and the ground-state vibrational coherence $\langle 0, 0|\rho|0, 1\rangle$ reflected by the nonvanishing coefficients $\mathcal{L}_{ii,12} = \sum_{\nu=3}^{6} \gamma n_{\omega_{2\nu}}^{T_1} (U_{4\nu} + U_{6\nu}) U_{\nu i}^T$, $\mathcal{L}_{jj,12} = \sum_{\nu=3}^{6} \gamma n_{\omega_{1\nu}}^{T_1} (U_{3\nu} + U_{5\nu}) U_{\nu j}^T$, $i = 3, 5; j = 4, 6$ in Eq. (7). This indeed breaks the secular approximation and results in the considerable enhancement of the population in pigment $B$ as shown in Fig. 4(a) (blue line exceeds much purple line). Thus the excitation energy transferred to pigment $B$ is considerably promoted. In contrast, it should be noted that the dynamics of excitation populations becomes decoupled to that of vibrational coherence without including the incoherent environment, namely $\mathcal{L}_{33,34}, \mathcal{L}_{33,43}, \mathcal{L}_{44,34}, \mathcal{L}_{44,43}, \mathcal{L}_{55,56}, \mathcal{L}_{55,65}, \mathcal{L}_{66,56}, \mathcal{L}_{66,65} = 0$, based on the structure of QME Eq. (7) and (S20) in SI. In this case with low-energy noise from protein included only, our results in Fig. 4(b) show that neither excited-state vibrational nor ground-state vibrational coherence can affect much the excitation energy transfer from pigment *A* to *B*. These analyses further elucidate the importance of considering the effect of incoherent radiation in addition to the low-energy fluctuations for rendering the long-lived ground-state vibrational coherence to enhance the excitation energy transfer. This in fact, is originated from the nonequilibriumness of the system, which will be discussed later on in the paper. Furthermore, Fig. 4(a) shows that the cumulative population on pigment *B*: $\int \overline{P}_B dt$ is much larger than the one without including exciton-vibrational coupling. This means that the total energy transport is much enhanced by the vibrational coherence, when including the radiations. This promotion of energy transport, in fact, is natural from the view of point of nonequilibriumness, which will be illustrated later.

**Nonequilibriumness, steady-state coherence and energy transfer.** It is important to note that the nonvanishing steady-state coherence is reached for both of the localized and delocalized cases, as shown in Fig. 3. To elucidate this in detail, we explore the nonequilibrium effect (induced by two heat sources, one is from radiations and the other is from phonons) on steady-state coherence, as shown in Fig. 4(d). It shows that both electronic and excitonic coherences are enhanced by decreasing the temperature of low-frequency fluctuations (phonons). This with the fixing temperature of incoherent radiations effectively increases the degree of nonequilibriumness characterized by detailed-balance-breaking. It indicates that the enhancement of steady-state quantum coherence can be attributed to the time-irreversibility (from nonequilibriumness by detailed-balance-breaking) of the whole system. It is also demonstrated that the steady-state coherence is considerably promoted in the far-from-equilibrium regime[37,38].

Moreover, the promotion of energy transfer discussed above can be understood from the underlying nonequilibrium feature with the detailed-balance-breaking at steady state as we suggested[35,37–40], since the nonequilibriumness generated by two heat sources can funnel the path and subsequently facilitate the unidirectional energy transfer, as shown in Fig. 4(c). As pointed out above, the breakdown of time-reversal symmetry at steady state from the violation of the detailed balance plays an essential role for long-survived oscillation of coherence to assist the enhancement of energy transfer.

On the other hand, the general mechanism uncovered in our model demonstrates that the slowing-down of dynamical decoherence is dominated by the suppression of exciton-environment interaction, rather than the time-irreversibility. Hence from the Lindblad equation, it is important to emphasize that the typical time scale for decoherence is mainly governed by the system-environment interaction, while the time-irrversibility from the detailed-balance-breaking is mostly responsible for efficient energy transport and the improvement of coherence at steady state[37,38,40], shown in Fig. 4(c).

**Effect of multiple vibrational modes.** It is worthy to point out that multiple vibrational modes will considerably elongate the surviving time of excitonic coherence, as generally elucidated in our model. As shown in Fig. S1 in SI, the electronic coherence by including two vibrational modes survives much longer than that by including one mode only (Fig. 3). Quantitatively $\overline{\tau}_2/\overline{\tau}_1 \sim 10$, which agrees with our theoretical estimation at the beginning.

## Discussion and Conclusion

Our work uncovers a mechanism in a general scenario for the long-lived coherence while considering the exciton-vibration interactions. The bare exciton is surrounded by a cloud consisting of discrete vibrational modes. This forms a new composite called polaron. The interactions between the system and the environments are consequently suppressed with respect to that of the bare excitons. This suggests that vibrational coherences generated by exciton-vibron coupling play a significant role in improving greatly the surviving time of the electronic and the excitonic coherences. This general mechanism has never been uncovered before to the best of our knowledge, although some investigations show the existence of the long-lived coherence, at numerical level[18,24]. Besides, the role of the ground-state coherence was also uncovered, which was elucidated to be non-trivial and essential for promoting the excitation energy transfer in pigment-protein complexes. The approaches popularly applied before were also shown to fail in predicting the role of vibrational coherence, especially the ground-state coherence on dynamical excitonic energy transfer. We also illustrated that the nonvanishing steady-state coherence is promoted by the time-irreversibility originating from the detailed-balance-breaking of the system. Furthermore our results on the slowing-down of dynamical decoherence from weaker coupling to environment and nonvanishing steady-state coherence from detailed-balance-breaking leading to efficient energy transfer provide a possible way to optimize the quantum information process.

## Acknowledgements

We acknowledge the support from the grant NSF-MCB-0947767 and thank Prof. Thomas Bergeman for polishing the writing.

## Author Contributions

Z. Zhang and J. Wang conceived the idea. Z. Zhang carried out the derivation and calculations. All authors contributed to the discussion of the project and writing of the manuscript.

## Additional Information

**Supplementary information** accompanies this paper at http://www.nature.com/srep

**Competing financial interests:** The authors declare no competing financial interests.

**How to cite this article**: Zhang, Z. and Wang, J. Origin of long-lived quantum coherence and excitation dynamics in pigment-protein complexes. *Sci. Rep.* **6**, 37629; doi: 10.1038/srep37629 (2016).